\documentclass[useAMS,usenatbib]{mn2e}
\usepackage{graphicx}%for EPS, PS files

\def\araa{ARA\&A}%
\def\apj{ApJ}%
\def\apjl{ApJ}%
\def\aap{A\&A}%
\def\cqg{Class. Quantum Grav.}%
\def\mnras{MNRAS}%
\def\ptrsa{Phil. Trans.:Phys. Sc. \& Eng.}
\def\prd{Phys.~Rev.~D}%
\def\prl{Phys.~Rev.~Lett.}%
\def\pasp{PASP}%
\def\sovast{Soviet~Ast.}%
\def\nat{Nature}%
\footnotesize
\newdimen\digitwidth    %define ! a one digit width for tables
\setbox0=\hbox{\rm0}
\digitwidth=\wd0
\catcode`!=\active
\def!{\kern\digitwidth}
\normalsize

\newcommand{\psr}{PSR~J1012$+$5307}

\title[Dipole radiation and grav. constant variation]{Generic tests of
  the existence of the gravitational dipole radiation and the
  variation of the gravitational constant} \author[K.Lazaridis et al.]
{ K.~Lazaridis,$^1$\thanks{Member of the International Max Planck
    Research School (IMPRS) for Astronomy and Astrophysics at the
    Universities of Bonn and Cologne} N.~Wex,$^1$A.~Jessner,$^1$
  M.~Kramer,$^1$$^,$$^2$ B.~W.~Stappers,$^2$$^,$$^3$ \newauthor \
  G.~H.~Janssen,$^4$ G.~Desvignes,$^5$ M.~B.~Purver,$^{2}$
  I.~Cognard,$^5$ G.~Theureau,$^5$ \newauthor \ A.~G.~Lyne,$^{2}$
  C.~A.~Jordan,$^{2}$ J.~A.~Zensus$^1$
  \\
  $^{1}$Max-Planck-Institut f\"ur Radioastronomie, Auf dem H\"ugel 69,
  53121, Bonn, Germany
  \\
  $^{2}$University of Manchester, Jodrell Bank Centre for
  Astrophysics, Alan Turing Building, Manchester M13 9PL, UK
  \\
  $^{3}$Stichting ASTRON, Postbus 2, 7990 AA, Dwingeloo, the
  Netherlands
  \\
  $^{4}$Astronomical Institute Anton Pannekoek, University of
  Amsterdam, Postbus 94249, 1090 GE Amsterdam, The Netherlands
  \\
  $^{5}$Laboratoire de Physique et Chimie de l'Environnement, CNRS, 3A
  Avenue de la Recherche Scientifique, \\ $\,$ 45071 Orl\'{e}ans Cedex
  2, France }
\let\leqslant=\leq %Authors with AMS fonts and mssymb.tex can comment
\date{}
\begin{document}

\maketitle
\newcommand{\setthebls}{
%                 de-comment this line for double spacing:
 %\baselineskip=20pt
}

\setthebls

\begin{abstract} 
  We present results from the high precision timing analysis of the
  pulsar-white dwarf (WD) binary \psr\ using 15 years of
  multi-telescope data. Observations were performed regularly by the
  European Pulsar Timing Array (EPTA) network, consisting of
  Effelsberg, Jodrell Bank, Westerbork and Nan\c cay. All the timing
  parameters have been improved from the previously published values,
  most by an order of magnitude. In addition, a parallax measurement
  of $\pi = 1.2(3)$\,mas is obtained for the first time for \psr,
  being consistent with the optical estimation from the WD
  companion. Combining improved 3D velocity information and models for
  the Galactic potential the complete evolutionary Galactic path of
  the system is obtained. A new intrinsic eccentricity upper limit of
  $e<8.4\times 10^{-7}$ is acquired, one of the smallest calculated
  for a binary system and a measurement of the variation of the
  projected semi-major axis also constrains the system's orbital
  orientation for the first time. It is shown that \psr\ is an ideal
  laboratory for testing alternative theories of gravity. The
  measurement of the change of the orbital period of the system of
  $\dot{P}_{b} = 5(1)\times 10^{-14}$ is used to set an upper limit on
  the dipole gravitational wave emission that is valid for a wide
  class of alternative theories of gravity. Moreover, it is shown that
  in combination with other binary pulsars \psr\ is an ideal system to
  provide self-consistent, generic limits, based only on millisecond
  pulsar data, for the dipole radiation and the variation of the
  gravitational constant $\dot{G}$.
\end{abstract}


%%%%%%%%%%%%%%%%%%%%%%%%%%%%%%%%%%%%%%%%%%%%%%%%%%%%%%%%%%%%%%%%%%%%%%%%%%%%%%%%

\section{Introduction}
\label{sec:Intro}

\psr\ is a 5.3\,ms pulsar in a binary system with orbital period of
14.5\,h and a low mass companion 
\citep{nll+95}. It was discovered during a survey for short period pulsars
with the 76\,m Lovell radio telescope at Jodrell Bank. \cite{llfn95} reported
optical observations revealing an optical counterpart within $0.2\pm
0.5$\,arcsec of the pulsar timing position being consistent with a helium
white dwarf (WD) companion.

The optical observations of the WD companion provide unique
information about the evolution of the binary system and the radio
pulsar itself, such as comparing the cooling age of the companion with
the spin-down age of the pulsar \citep{llfn95,dsbh98,esg01}.

Using the NE2001 model for the Galactic distribution of free electrons
\citep{cl02} and the pulsar's dispersion measure (DM) of
$9$\,cm$^{-3}$ pc \citep{nll+95} a distance of $\sim 410$\,pc is
derived. In contrast, \cite{cgk98} compared the measured optical
luminosity of the WD to the value expected from WD models and
calculated a distance of $d = 840 \pm 90$\,pc. In addition they
measured, by the Doppler shift of the measured H spectrum of the
companion, a radial velocity component of $44 \pm 8$\,km\,s$^{-1}$
relative to the SSB. From the radial velocity and the orbital
parameters of the system the mass ratio of the pulsar and its
companion was measured to be $q=m_{p}/m_{c} = 10.5\pm 0.5$. Finally by
fitting the spectrum of the WD to a grid of DA (hydrogen dominated)
model atmospheres they derived a companion mass of $m_{c} = 0.16 \pm
0.02\,M_{\odot}$, a pulsar mass of $m_{p} = 1.64\pm 0.22\,M_{\odot}$
and an orbital inclination angle of $i=52^\circ \pm 4^\circ$.

\cite{lcw+01} presented the most complete precision timing analysis of
\psr\ using 4 years of timing data from the Effelsberg 100\,m radio
telescope and 7 years from the 76\,m Lovell telescope. Using their low
eccentricity binary model ELL1 and combining the timing measurements
with the results from the optical observations they derived the full
3D velocity information for the system. Furthermore, after correcting
for Doppler effects, they derived the intrinsic spin parameters of the
pulsar and a characteristic age of $8.6\pm 1.9$\,Gyr which is
consistent with the WD age from the optical estimates. In addition,
after calculating upper limits for an extremely low orbital
eccentricity they discussed evolutionary scenarios for the binary
system but also presented tests and limits of alternative theories of
gravitation. Finally, they discussed the prospects of future
measurements of Post-Keplerian parameters (PK) which can contribute to
the description of the orientation of the system and the calculation
of stringent limits for the effective coupling strength of the scalar
field to the pulsar.

In this paper we revisit \psr\ with seven more years of high-precision
timing data and combined datasets from the European Pulsar Timing
Array (EPTA) telescopes consisting of the 100\,m Effelsberg
radio-telescope of the Max-Planck-Institute for Radioastronomy,
Germany, the 76\,m Lovell radio-telescope at Jodrell Bank observatory
of the University of Manchester, UK, the 94\,m equivalent Westerbork
Synthesis Radio Telescope (WSRT), the Netherlands and the 94\,m
equivalent Nan\c cay decimetric Radio Telescope (NRT), France. After a
short description of the timing procedure and the technique of
combining our multi-telescope data, we present the updated
measurements of the astrometric, spin and binary parameters for the
system. Specifically we show the improvement in all the timing
parameters and in the orbital eccentricity limit and in addition the
value for the first time for \psr\ of the timing
parallax. Furthermore, we obtain a value for the orbital period
variation, in agreement with the prediction of \citep{lcw+01}, from
which we test different theories of gravitation and give one of the
tightest bounds on a Parametrised Post-Newtonian (PPN)
parameter. Finally, we present how the timing measurement of the
change of the projected semi-major axis can complete the picture of
the orientation of the binary system.

%%%%%%%%%%%%%%%%%%%%%%%%%%%%%%%%%%%%%%%%%%%%%%%%%%%%%%%%%%%%%%%%%%%%%%%%%%%%%%%%

\section{Observations}
\label{sec:obs}

\subsection{Effelsberg}
\label{subsec:Eff}

\psr\ was observed regularly with the Effelsberg 100\,m radio
telescope since October 1996 with typical observing times of
5--15\,min in three consecutive scans. Monthly observations were
performed at 1400\,MHz using the primary focus cooled HEMT
receiver. It has a typical system temperature of 25\,K and an antenna
gain of 1.5\,KJy$^{-1}$. In order to monitor dispersion measure (DM)
variations, it was also observed irregularly until August 2006 and
monthly thereafter, at 2700\,MHz. At these frequencies a cooled HEMT
receiver located at the secondary focus was used which has a system
temperature of 25\,K. Finally, it was occasionally observed at
860\,MHz using an uncooled HEMT receiver, located at the primary
focus, with a typical system temperature of 60\,K. The
Effelsberg-Berkeley Pulsar Processor (EBPP) was used for coherent
on-line de-dispersion of the signal from the LHC and RHC
polarisations. It has 32 channels for both polarisations spread across
bandwidths of 40, 100 and 80\,MHz at 860, 1400 and 2700\,MHz
respectively \citep{bdz+97}. The output signals of each channel were
fed into de-disperser boards for coherent on-line de-dispersion and
were synchronously folded with the topocentric period.

Each TOA was obtained by cross-correlation of the profile with a
synthetic template, which was constructed out of 12 Gaussian
components fitted to a high signal-to-noise ratio standard profile
\citep{kxl+98, kll+99}. The TOAs were locally time stamped using a
H-maser clock at the observatory. They were converted to UTC using
the GPS maser offset values measured at the observatory, and the GPS
to UTC corrections were made from the Bureau International des Poids
et Measures (BIPM\footnote{http://www.bipm.org}).

%%%%%%%%%%

\subsection{Jodrell Bank}
\label{sebsec:JBO}

\psr\ has been observed with the Lovell radio telescope 2--3 times per
month since 1993, at three different frequencies. It is continuously
observed at 1400\,MHz and at 410 and 606\,MHz, it was observed until
1997 and 1999, respectively. All the receivers are cryogenically
cooled with system temperatures of 25, 50 and 35, respectively and
their LHC and RHC polarisation signals are detected and incoherently
de-dispersed in a 2$\times $32$\times $0.0312\,MHz filter bank at
410\,MHz, in a 2$\times $6$\times $0.1250\,MHz filter-bank at 606\,MHz
and in a 2$\times $32$\times $1\,MHz filter-bank at 1400\,MHz.  The
signals are synchronously folded at the topocentric pulsar
period and finally copied to a disc.

Each TOA was obtained by cross-correlation of the profile with a
standard template, generated by the summation of high S/N
profiles. The TOAs were transferred to GPS from a H-maser and the
time stamp was derived as for Effelsberg.

%%%%%%%%%%

\subsection{Westerbork}
\label{subsec:WSRT}

\psr\ was observed monthly using the WSRT with the PuMa-I pulsar
machine \citep{vkh+02}. We used three observing frequencies:
observations at centre frequencies of 1380\,MHz and 350\,MHz were
carried out monthly from August 1999, and the pulsar was observed
occasionally at a centre frequency of 840\,MHz from 2000 until
2002. The system temperatures were 27, 120 and 75\,K, respectively and
most observations were 30 minutes long.  The WSRT observations used a
bandwidth of 8$\times$10\,MHz for observations at 840\,MHz and
1380\,MHz, and after September 2006 the 8 bands were spread out over a
total observing bandwidth of 160\,MHz for the 1380\,MHz
observations. The observations at the low frequency setup used only
two bands of 10\,MHz, either centred at 328 and 382\,MHz or 323 and
367\,MHz.  For the observations taken at 1380 or 840\,MHz we used 64
frequency channels per 10\,MHz band, and the observations at the low
frequencies used 256 frequency channels per 10\,MHz band.

For each observation, the data were de-dispersed and folded
offline. Integration over frequency and time resulted in one single
profile for each observation. Each profile was cross-correlated with a
standard template, generated by the summation of high S/N profiles, so
finally only one TOA was computed for each observation. The TOAs were
transferred to GPS from a H-maser clock and the time stamp was derived
as for Effelsberg.

%%%%%%%%%%

\subsection{Nan\c cay}
\label{subsec:NCY}

\psr\ was observed roughly every 3 to 4 weeks with the Nan\c cay Radio
Telescope (NRT) since late 2004. The Nan\c cay Radio Telescope is
equivalent to a 94-m dish, with a gain of 1.4\,K\,Jy$^{-1}$ and a
minimal system temperature of 35\,K at 1.4\,GHz in the direction of
the pulsar. With the BON (Berkeley-Orleans-Nan\c cay) coherent
dedispersor, in the period covered by the observations, a 64\,MHz band
centred on 1398 MHz is split into sixteen 4\,MHz channels and
coherently dedispersed using a PC-cluster, with typical integration
times of one hour. The Nan\c cay data are recorded on a UTC(GPS) time
scale marked at the analogue to digital converter by a Thunderbolt
receiver (Trimble Inc.). Differences between UTC and UTC(GPS) are less
than 10\,ns and therefore no laboratory clock corrections are
needed. A single TOA was calculated from a cross-correlation with a
pulse template for each observation of one hour.

%%%%%%%%%%

\subsection{Multi-telescope precision timing}
\label{subsec:multi}

Combing the EPTA multi-telescope datasets is not a trivial
process. The main technique for achieving the optimal combination of
the data sets is presented by \cite{jsk+08}. In general, using
different datasets from different telescopes and obtained at different
frequencies requires extra corrections, apart from the usual one of
the transformation of all the individual telescope arrival times to
arrival times in the TAI at the solar system barycentre (SSB). The
extra corrections needed are usually constant time offsets between
different datasets of residuals. These offsets derive from
differences in the procedure of calculating the TOAs at each
telescope, specifically differences in the templates. The timing
software package
TEMPO\footnote{http://www.atnf.csiro.au/research/pulsar/tempo/} can
fit for these time offsets or "jumps". In the current work seven
of these jumps need to be fitted corresponding not only to the
telescopes but also to the different frequencies used. Normally, three
"jumps", one for each telescope, would be sufficient. However, the
TOAs at different frequencies are usually calculated by
different templates, which might not be aligned optimally. In the
current case, this occurs for Effelsberg and Westerbork TOAs. In
Table \ref{tab:telescopes} the properties of the individual datasets
are presented.

%-------------------------------------------------------------------------------

\begin{table*}
\center
\caption{Properties of the individual telescope datasets.}
\begin{tabular}{lcccc}
\hline
%\hline
    {Properties}     &{Effelsberg} & {Jodrell Bank}  & {Westerbork} & {Nan\c cay} \\
\noalign{\smallskip}
\hline
\noalign{\smallskip}
Number of TOAs                 &  1972        & 600          & 234          & 86 \\
Time span (MJD)                & 50371--54717 & 49221--54688 & 51389--54638 & 53309--54587 \\
Rms of individual set($\mu s$) &  2.7         & 8.6          & 2.9          & 1.9 \\
Observed frequencies (MHz)     &  860, 1400, 2700 & 410, 606, 1400 & 330, 370, 840, 1380 & 1400\\
\noalign{\smallskip}
\hline
\end{tabular}
\label{tab:telescopes}
\end{table*}

%-------------------------------------------------------------------------------

The combination of the EPTA datasets has many advantages
\citep{jsk+08}. The need for continuous multi-frequency TOAs for
precise measurement of the dispersion measure (DM) and monitoring of
the DM variations was successfully accomplished.
% The obtained upper limit for DM variations is $\dot{\rm DM} <
% 2.6\times 10^{-5}$\,cm$^{-3}$ pc\,yr$^{-1}$.
Most important the combination of the high quality data
from Effelsberg, Nan\c cay, WSRT with the long time span data of
Jodrell (\& Effelsberg) provides us with a 15 year dataset of TOAs
with no significant time gaps. Using all these EPTA datasets we
improve and measure all the astrometric, spin and binary parameters of
\psr\ presented in the first column of Table \ref{tab:par}. As a
comparison, in the second column the measured parameters of only the
current Effelsberg set is shown and in the third the Effelsberg
measurements from \cite{lcw+01}. From Table \ref{tab:par} it is clear
that the EPTA provides the most accurate error estimations and in
addition $\sim 3\sigma$ measurements of 2 post-Keplerian parameters.

%-------------------------------------------------------------------------------

\begin{table*}
\center
\caption{Timing parameters for \psr. Comparison between the datasets from EPTA, 
Effelsberg until 2008 and Effelsberg until 2001 \citep{lcw+01}.}
\begin{tabular}{lccc}
\hline
%\hline
{Parameters} &EPTA & {Effelsberg 2008}    & {Effelsberg 2001} \\
\noalign{\smallskip}
\hline
\noalign{\smallskip}
Right ascension, $\alpha$ (J2000) & 10$^h$12$^m$33$^s$.4341010(99) & 10$^h$12$^m$33$^s$.434089(13) & 10$^h$12$^m$33$^s$.43364(3) \\
Declination,     $\delta$ (J2000) & 53$^\circ$07'02''.60070(13) &  53$^\circ$07'02".6001(2) & 53$^\circ$07'02''.5878(4) \\ 
$\mu_{\alpha}$ (mas\,yr$^{-1})$   & 2.562(14)   & 2.56(2)     & 2.62(13)   \\
$\mu_{\delta}$ (mas\,yr$^{-1})$   & $-$25.61(2) & $-$25.49(2) & $-$25.0(2) \\
Parallax, $\pi$ (mas)             &    1.22(26) &      0.8(3) &  $<1.6$    \\ 
\\
$\nu$ (Hz)              & 190.2678376220576(5)           &    190.2678376220611(8)        & 190.267837621910(3)           \\
$\dot{\nu}$ (s$^{-2}$)  & $-$6.20063(3)$\times 10^{-16}$ & $-$6.20077(5)$\times 10^{-16}$ & $-$6.2070(5)$\times 10^{-16}$ \\
$P$ (ms)                & 5.255749014115410(15)          & 5.25574901411531(2)            & 0.00525574901411947(7)        \\ 
$\dot{P}$ (s\,s$^{-1}$) &  1.712794(9)$\times 10^{-20}$  & 1.712833(13)$\times 10^{-20}$  & 1.71456(15)$\times 10^{-20}$  \\  
\\
Epoch (MJD) & 50700.0 & 50700.0 & 50700.0 \\ 
\\ 
Dispersion measure, DM (cm$^{-3}$ pc) & 9.02314(7) & 9.0209(3) & 9.022(3) \\ 
\\ 
Orbital period, $P_{b}$ (days)  & 0.60467271355(3)& 0.60467271355(4)& 0.6046727133(2) \\ 
Projected semi-major axis, $x$ (lt-s) & 0.5818172(2) & 0.5818175(2)& 0.5818174(5) \\ 
$\eta$   ($\equiv e\sin\omega$) & 1.2(3)$\times 10^{-6}$ & 1.6(3)$\times 10^{-6}$ & 1.1(5)$\times 10^{-6}$ \\ 
$\kappa$ ($\equiv e\cos\omega$) & 0.06(31)$\times 10^{-6}$ & 0.14(34)$\times 10^{-6}$ & 0.20(50)$\times 10^{-6}$ \\
Eccentricity, $e$ $^\ast$ & 1.2(3)$\times 10^{-6}$ & 1.6(3)$\times 10^{-6}$ & 1.1(5)$\times 10^{-6}$ \\
Longitude of the periastron, $\omega$ (deg) $^\ast$ & 93(14) & 85(12) & 79(24) \\
$T_{ASC}$ (MJD) & 50700.08162891(4) & 50700.08162891(5) & 50700.08162905(9) \\ 
\\ 
$ \dot{P_{b}}$ (s\,s$^{-1}$) & 5.0(1.4)$\times 10^{-14}$ & 4(2)$\times 10^{-14}$ & 0.3(3)$\times 10^{-12}$ \\ 
$ \dot{x}$     (s\,s$^{-1}$) &   2.3(8)$\times 10^{-15}$ & $<1.8\times 10^{-15}$ & $<1.8\times 10^{-15}$   \\ 
\\ 
Solar system ephemeris model & DE405 & DE405 & DE200 \\ 
Number of TOAs               & 2892  & 1972  & 1213  \\ 
RMS timing residual ($\mu$s) & 3.1   & 2.7   & 3.1   \\ 
\noalign{\smallskip} \hline
\multicolumn{4}{l}{\footnotesize{$^\ast$ The eccentricity and the longitude of
    the periastron are calculated from the Laplace-Lagrange parameters, $\eta$ 
    and $\kappa$.}} \\ 
\multicolumn{4}{l}{\footnotesize{Figures in parentheses are the nominal
    1$\sigma$ TEMPO uncertainties in the least-significant digits quoted}} \\
\end{tabular}
\label{tab:par}
\end{table*}

%%%%%%%%%%%%%%%%%%%%%%%%%%%%%%%%%%%%%%%%%%%%%%%%%%%%%%%%%%%%%%%%%%%%%%%%%%%%%%%%

\section{Analysis \& Results}
\label{sec:analysis}

\begin{figure}
\begin{center}
  \mbox{\includegraphics[width=0.53\textwidth]{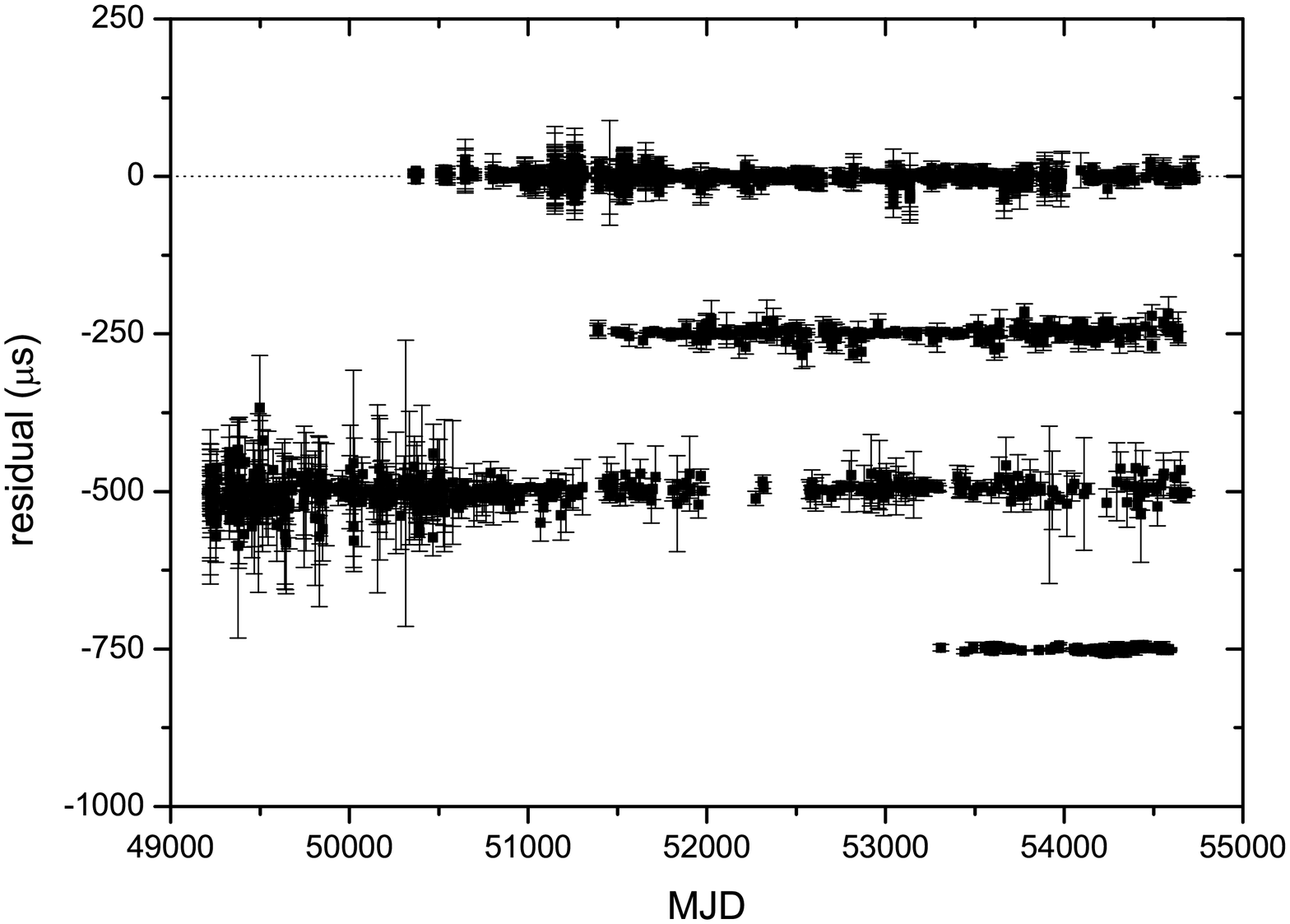}}
  \mbox{\includegraphics[width=0.53\textwidth]{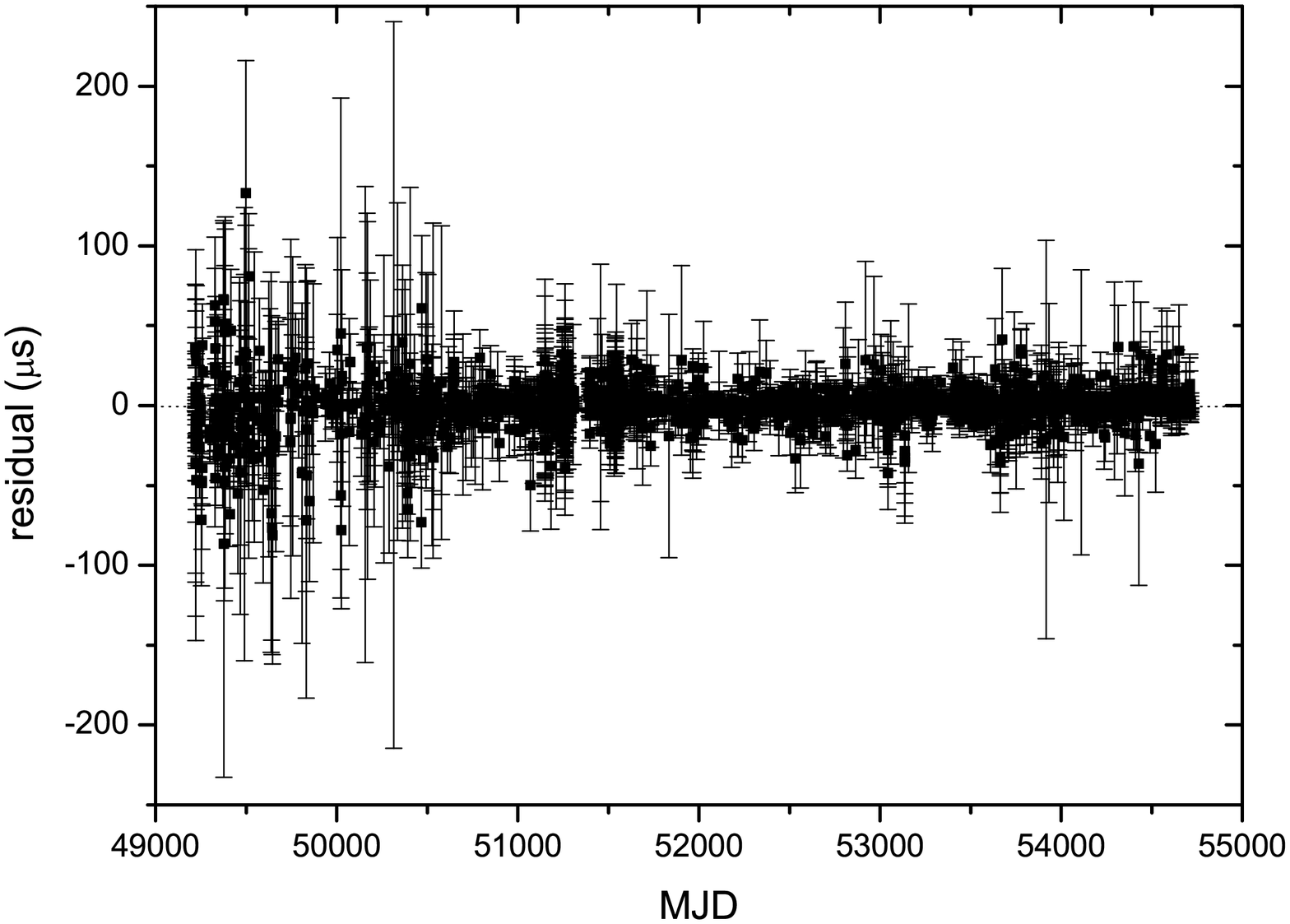}}
  \caption{(Top) Post-fit timing residuals for the dataset of each
    telescope. From top to bottom, Effelsberg, Westerbork, Jodrell and
    Nan\c cay. (Bottom) Post-fit timing residuals. Best timing
    solution with all the data sets yields the parameters in
    Table \ref{tab:par}.}
\label{fig:postfit}
\end{center}
\end{figure}

All the combined TOAs, weighted by their individual uncertainties,
were analysed with TEMPO, using the DE405 ephemeris of the Jet
Propulsion Laboratory (JPL) \citep{sta98b, sta04b} and the ELL1
\citep{lcw+01} binary model. TEMPO minimises the sum of the weighted
squared timing residuals, producing a set of improved pulsar
parameters and the post-fit timing residuals. The uncertainties on the
TOAs from each telescope are scaled by an appropriate factor to
achieve a uniform reduced $\chi ^2\simeq 1$ for each data set. The
best post-fit timing solution of all the combined residuals is
presented in Figure \ref{fig:postfit}. In the top panel the post-fit
versus time is shown, with arbitrary offsets of the different
datasets, where it is clear that the uncertainties of most of the data
points are comparable. 
%(apart from most Jodrell Bank which are a bit bigger and Nan\c cay which are smaller). 
By comparing the
parameters we get by different combinations of data sets
(i.e. only the 1400\,MHz data or without Jodrell) we concluded that it
is much more efficient to finally use all the available datasets
together, as shown in the lower panel of Figure \ref{fig:postfit}.

%-------------------------------------------------------------------------------
 
%%%%%%%%%%

\subsection{Timing parallax \& distance}
\label{subsec:parallax}

Apart from the common method of DM distance estimation ($\sim
410$\,pc) and the more rare optical one ($d=840\pm 90$\,pc) for \psr,
there is another way of measuring the distance to a pulsar, with
pulsar timing. In general, the timing residuals of nearby pulsars
demonstrate an annual parallax. This timing parallax is obtained by
measuring a time delay of the TOAs caused by the curvature of the
emitted wavefronts at different positions of the Earth in its
orbit. This time delay has an amplitude of
$r_{E\odot}^2\cos\beta/(2cd)$ \citep{lk05}, where $r_{E\odot}$ is the
Earth-Sun distance, $\beta$ the ecliptic latitude of the pulsar, $c$
the speed of light and $d$ the distance to the pulsar. This effect has
been measured for very few pulsars like PSR B1855+09 \citep{ktr94},
PSR J1713+0747 \citep{cfw94}, PSR J0437$-$4715 \citep{sbm+97}, PSR
J1744$-$1134 \citep{tsb+99}, PSR J2145$-$0750 \citep{lkd+04} and PSR
J0030+0451 \citep{lkn+06}. Here for the first time we measure a
parallax $\pi = 1.2\pm 0.3$\,mas for \psr. This parallax corresponds
to a distance of $d = 822 \pm 178$\,pc which is consistent with the $d
= 840 \pm 90$ pc measured from the optical observations. The
difference with the DM distance may point to a sparse free electron
distribution in this location of the Galaxy \citep{gmcm08, cbv+09}. By
combining the optical and timing parallax distance measurements we
calculate the weighted mean \citep{wj03} of the distance getting an
improved value of $d = 836 \pm 80$\,pc.

%%%%%%%%%%

\subsection{Improved 3D velocity measurement \& Galactic motion}
\label{subsec:velocity}

Combining the proper motion measured from timing (Table \ref{tab:par})
and the distance to the system and the radial velocity of $v_r = 44
\pm 8$ km\,$^{-1}$s from the optical observations of the WD,
\cite{lcw+01} managed to determine the full 3D motion of the pulsar
relative to the SSB. Our new timing results improve the proper motion
measurements by an order of magnitude and using the combined parallax
and optical distance we recalculate the 3D motion of the pulsar. We
derive transverse velocities of
\begin{equation}
\label{eq:velocity1}
v_{\alpha} = \mu_{\alpha} d = 10.2 \pm 1.0\,{\rm km\,s^{-1}}
\end{equation}
and
\begin{equation}
\label{eq:velocity2}
v_{\delta} = \mu_{\delta}d = 101.5 \pm 9.7\,{\rm km\,s^{-1}}
\end{equation}
This yields a total transverse velocity of $v_t = 102.0\pm
9.8$\,km\,s$^{-1}$. Using the radial velocity from the optical
measurements we get the space velocity of the system $v_{space} =
111.4 \pm 9.5$\,km\,s$^{-1}$, consistent and almost three times more precise
than the previous value. In addition, this value is still
consistent with the average space velocity of millisecond pulsars of
$130$\,km\,s$^{-1}$ \citep{lml+98, tsb+99}.

\begin{figure}
\begin{center}
  \mbox{\includegraphics[width=0.54\textwidth]{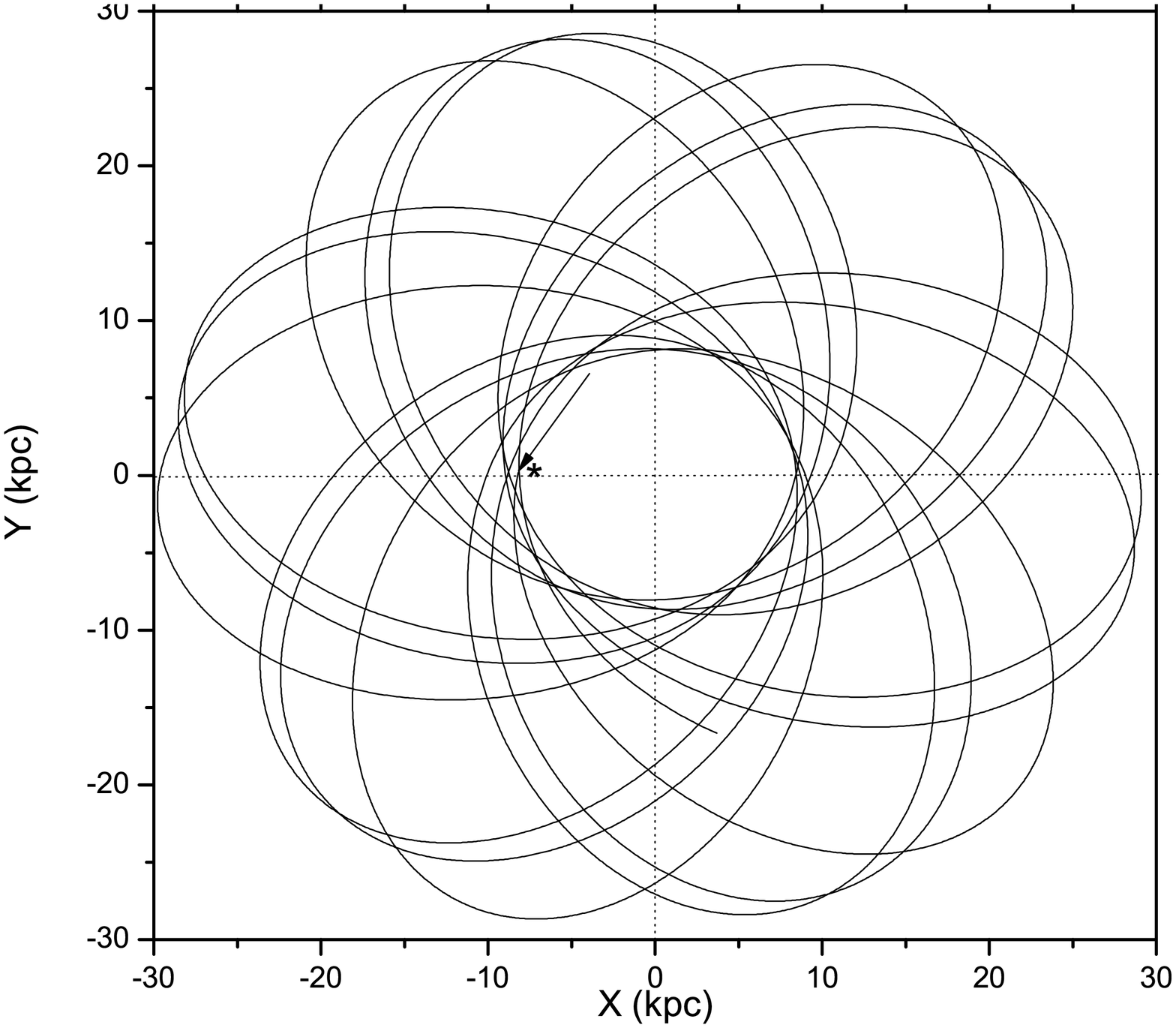}}
  \mbox{\includegraphics[width=0.54\textwidth]{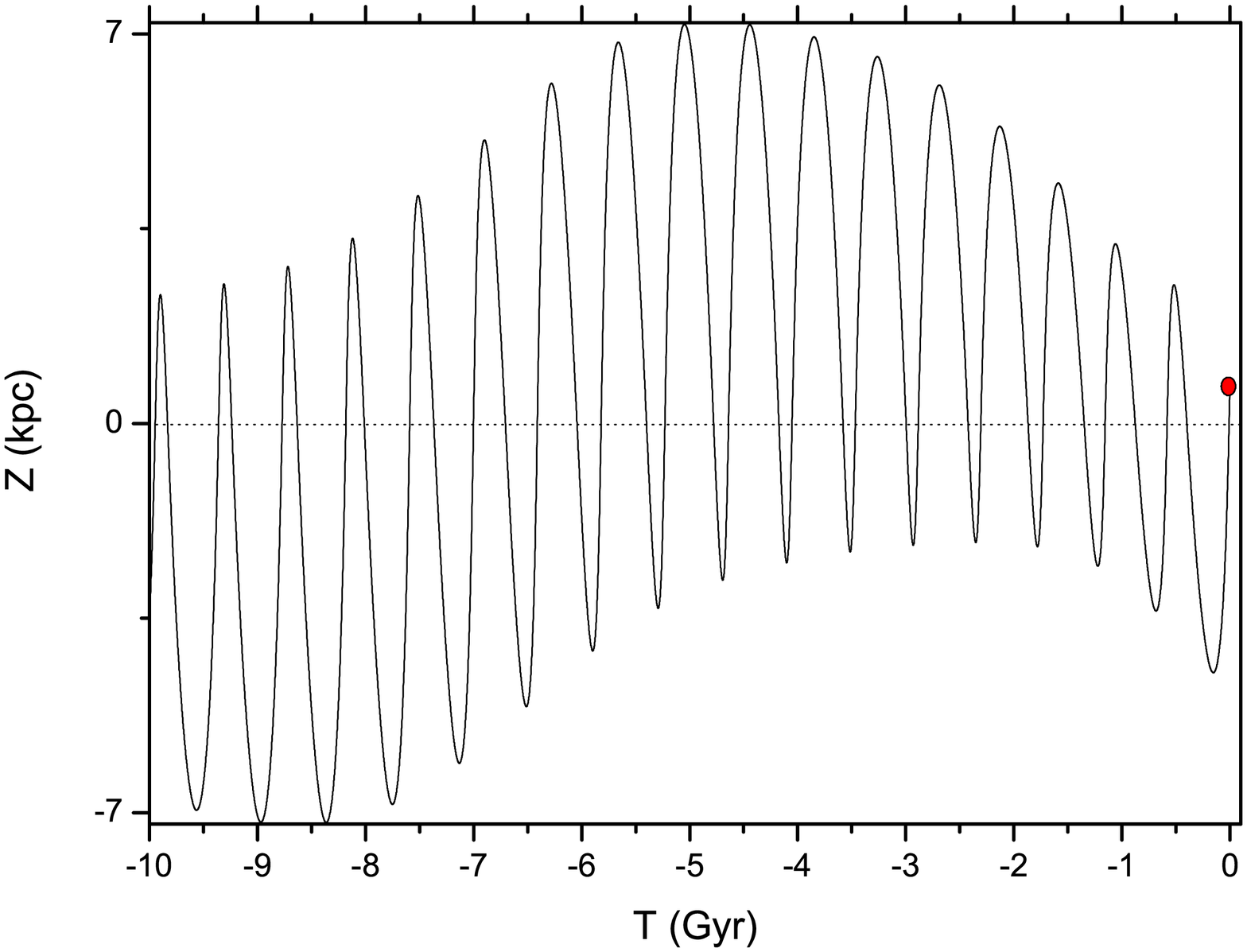}}
  \caption{(Top) Evolutionary path of \psr\ on the Galactic plane. With
    arrow the current position of the pulsar and with $\ast$ the
    position of the Sun is noted. (Bottom) The oscillations of \psr\
    above and below the Galactic plane through time.}
\label{fig:xyz}
\end{center}
\end{figure}

Since we know the 3D velocity of \psr\ we can try, for the first time,
to track its Galactic path in time and space. Assuming a
characteristic age of $\sim 10$\,Gyr and applying a model for the
Galactic potential \citep{kg89, pac90}, we derive the evolutionary
path of \psr\ in the Galaxy from the point it started emitting as a
millisecond pulsar. In the top part of Figure \ref{fig:xyz}, the
projection of the evolutionary path of the pulsar on the Galactic
plane is shown, where the arrow indicates the current position of the
pulsar and the star indicates the position of the Sun, for the
\cite{kg89} model (the \cite{pac90} derives similar results). It is
obvious that the pulsar is presently at one of its closest approaches
to the Sun, which is why we can actually observe it. \psr reaches
maximum distances of $\sim 30$\,kpc through its path, spending only a
small fraction of its lifetime close to the solar system orbit. In
Figure \ref{fig:xyz} (bottom) the movement of \psr\ above and below
the Galactic plane is shown versus time indicating that the pulsar is
oscillating with a period of $\sim 0.6$\,Gyr reaching a maximum
distance of $\sim 7$\,kpc above and below the Galactic plane.

%-------------------------------------------------------------------------------

%%%%%%%%%%

\subsection{Eccentricity}
\label{subsec:eccentricity}

\begin{figure}
\begin{center}
  \mbox{\includegraphics[width=0.43\textwidth]{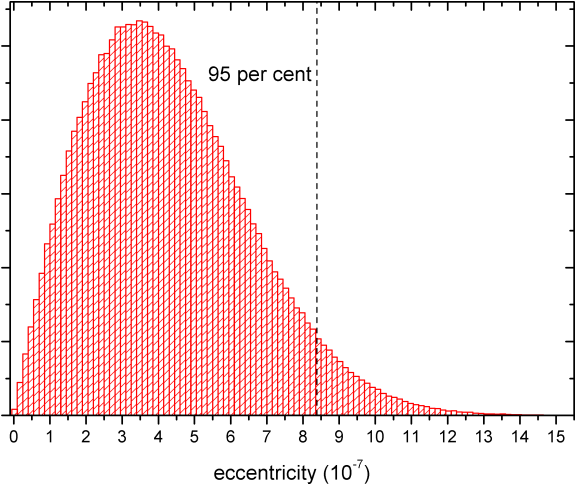}}
  \caption{Distribution of values for the intrinsic eccentricity from
    Monte Carlo simulations.The dashed line cuts the distribution at
    the 95 per cent of the values.}
\label{fig:MCecc}
\end{center}
\end{figure}

\psr\ is a low eccentricity binary system. In our current timing solution we
measure a value for the eccentricity of $(1.2 \pm 0.3) \times
10^{-6}$. However, as shown in \cite{lcw+01}, the Shapiro delay cannot be
separated from the Roemer delay for this system, which leads to a small
correction to this eccentricity value and specifically to the first
Laplace-Lagrange parameter $\eta = e\sin\omega$. Thus, following their
convention, for a companion mass of $m_{c} = 0.16(2)\, M_{\odot}$, a mass
ratio $q = 10.5(5)$ and a mass function of $f_{m} = 0.000578\,M_{\odot}$ we
derive the range $r$ and shape $s$ of the Shapiro delay according to
\begin{equation}
\label{eq:range}
r = 4.9255 \, (m_{c}/M_{\odot}) \, \mu{\rm s}
\end{equation}
and
\begin{equation}
\label{eq:shape}
s = \sin i = \left[\frac{f_{m}(q+1)^{2}}{m_{c}}\right]^{1/3}.
\end{equation}
The intrinsic value of $\eta$, calculated from equation (A22) of
\cite{lcw+01}, due to the contribution of the Shapiro delay, is
$\eta=(-1.4\pm 3.4)\times 10^{-7}$. The true eccentricity of the
system is $e=\sqrt{\eta^{2}+\kappa^{2}}$, where $\kappa=e\cos
\omega=(0.6\pm3.1)\times 10^{-7}$. By solving this equation in a Monte
Carlo simulation (Figure \ref{fig:MCecc}), for datasets of the values
and uncertainties of the intrinsic $\eta$ and $\kappa$, we obtain an
upper limit for the intrinsic eccentricity:
\begin{equation}
e < 5.2\times 10^{-7} \qquad (68\;{\rm per\;cent}\;{\rm C.L.})
\end{equation}
\begin{equation}
e < 8.4\times 10^{-7} \qquad (95\;{\rm per\;cent}\;{\rm C.L.})
\end{equation}
This limit is better than the previously published value \citep{lcw+01}.

This improved limit has another significant importance. Using the
fluctuation-dissipation theorem, \cite{phi92} predicted that the
orbital eccentricity, of a pulsar-WD binary system is correlated with
the orbital period. Specifically, there is the theoretical prediction
of a relic orbital eccentricity due to convective eddy currents in the
mass accretion process of the neutron star from the companion while in
the red giant phase. In \cite{lcw+01} the eccentricity limit of \psr\
was plotted versus the orbital period and was compared with the model
curves of the \cite{pk94} model. Our current eccentricity limit is
much lower than the one in \cite{lcw+01}, but still in good agreement
with the predictions from this model.

%%%%%%%%%%

\subsection{Changes in the projected semi-major axis}
\label{subsec:xdot}

A change in the projected semi-major axis has been measured in the current
analysis, for the first time, for PSR 1012+5307. The observed value of $\dot{x}
_{obs} = 2.3(8)\times 10^{-15}$ can be the result of the various effects 
\citep{lk05}:
\begin{equation}
\label{eq:dotA1}
\dot{x}_{obs} = \dot{x}^D + \dot{x}^{GW} + \frac{d\epsilon_A}{dt}+
                \dot{x}^{\dot{m}} + \dot{x}^{SO} + \dot{x}^{planet} + \dot{x}^{PM}.
\end{equation}

The first term, $\dot{x}^D$, is the Doppler correction, which is the combined
effect of the proper motion of the system \citep{shk70} and a correction term
for the Galactic acceleration. The contribution for the Galactic acceleration,
$\dot{x}^{Gal}$, is of order $6 \times 10^{-20}$.  Furthermore, we calculate
the contribution of the Shklovskii effect to be $\dot{x}^{Shk} = x
(\mu_{\alpha}^{2} + \mu_{\delta}^{2}) d / c \sim 8 \times 10^{-19}$. Both the
contributions are very small compared to the observed value, thus, this term
can be neglected.

The second term, $\dot{x}^{GW}$ is arising from the shrinking of the
orbit due to gravitational-wave damping 
\begin{equation}
\label{eq:xdotGR}
\dot{x}^{GW} = -x \,\frac{64}{5}\left(\frac{2\pi}{P_b}\right)^{8/3}
               \frac{(T_\odot m_c)^{5/3}q}{(q+1)^{1/3}}
             = (-8.2 \pm 1.7) \times 10^{-20}
\end{equation}
\citep{pet64}, where $T_\odot = GM_\odot/c^3 = 4.9255\,\mu s$ and $m_c$ is
expressed in units of solar masses. $x$ is the projected semi-major axis and
$P_{b}$ the orbital period. This contribution again is much smaller than the
current measurement precision.

The third term, $d\epsilon_{A}/dt$, is the contribution of the varying
aberration caused by geodetic precession of the pulsar spin axis, and is
typically of order $\Omega^{geod}P/P_b \approx 2\times 10^{-18}$
\citep{dt92}. For a recycled pulsar, like \psr, the spin is expected to be
close to parallel to the orbital angular momentum, which further suppresses
this effect. Hence, the contribution is at least three orders of magnitude
smaller than the observed.

The fourth term, $\dot{x}^{\dot{m}}$, is representing a change in the
size of the orbit caused by mass loss from the binary system. We
investigate the mass loss due to the loss of rotation energy
by the pulsar, which we consider as the dominant mass-loss effect. We
initially calculate the change in the orbital period from the same
contribution as follows:
\begin{equation}
\label{eq:massloss}
\dot{P}_{b}^{\dot{m}} = 8\pi^2 \, 
  \frac{I_{p}}{c^2M} \, \frac{\dot{P}}{P^3}\,P_b \sim 10^{-16} , 
\end{equation}
where $M = m_p + m_c$ and the moment of inertia of the pulsar $I_{p}\sim
10^{45}$ g\,cm$^{2}$. Subsequently, by Kepler's third law we calculate the
change in the projected semi-major axis of the orbit to be $\sim
10^{-17}$. Thus, we can also neglect this contribution.

The fifth and the sixth terms, $\dot{x}^{SO}$ and $\dot{x}^{planet}$, are the
contributions due to the classical spin-orbit coupling caused by a
spin-induced quadrupole moment of the companion and the existence of an
additional planetary companion respectively. They can both be neglected. For
the first one to be significant, a main-sequence star or a rapidly rotating
white dwarf companion \citep{wjm+98, klm+00} would be necessary. The second is
not being considered because there is no evidence for another companion to the
pulsar.

\begin{figure}
\begin{center}
  \mbox{\includegraphics[width=0.52\textwidth]{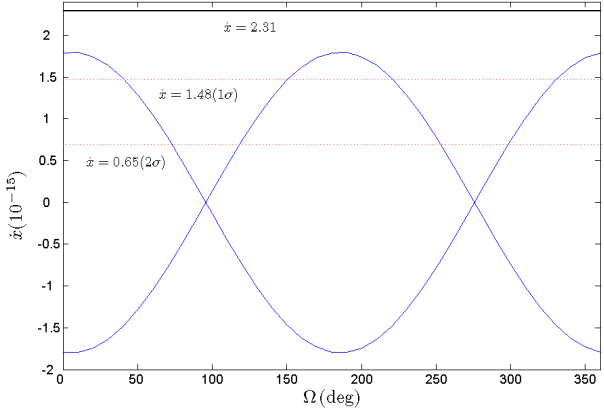}}
  \caption{Change of the projected semi-major axis versus position
    angle of the ascending node. The two curves have been produced for
    $i=52^\circ$ (peak at $180^\circ$) and $i=128^\circ$ (peak at
    $0^\circ$). The solid line represents the measured value of
    $\dot{x}$ and the dashed lines the 1$\sigma$ and 2$\sigma$ limits
    of $\dot{x}$, from top to bottom. The latter constrains the
    orientation $\Omega$.}
\label{fig:XdotOmega}
\end{center}
\end{figure}

Since all the other contributions are much smaller than the observed
variation of the projected semi-major axis, we conclude that the
measured value is arising from the last term of equation
(\ref{eq:dotA1}), $\dot{x}^{PM}$. This is a variation of $x$ caused by
a change of the orbital inclination while the binary system is moving
relatively to the SSB \citep{ajrt96, kop96, sbm+97}. The measurement
of the effect is presented in the following equation:
\begin{equation}
\label{eq:dotx}
 \dot{x}^{PM} = 1.54\times 10^{-16} \, x \, \cot i \,
                (-\mu_{\alpha}\sin\Omega + \mu_{\delta}\cos\Omega) \:,
\end{equation}
where $\Omega$ is the position angle of the ascending node. The quantities
$x$, $\mu_\alpha$ and $\mu_\delta$ are expressed in seconds and
milliarcseconds per year, respectively. The proper motions and the inclination
angle have been measured and since we measure the value of
$\dot{x}^{PM}=\dot{x}_{obs}$, we can, for the first time, restrict the orbital
orientation $\Omega$ of \psr. In Figure \ref{fig:XdotOmega} the $\dot{x}^{PM}$
versus the position angle of the ascending node is presented. Unfortunately,
our measured value cannot fully restrict the orientation, however from the
lower limits of $\dot{x}_{obs}$ we derive significant limits for the
position angle. For an inclination angle of $i=52^\circ$ we get
\begin{equation}
151^\circ < \Omega < 220^\circ \qquad (68\;{\rm per\;cent}\;{\rm C.L.})
\end{equation}
and
\begin{equation}
117^\circ < \Omega < 255^\circ \qquad (95\;{\rm per\;cent}\;{\rm C.L.}) ,
\end{equation}
while for $i=128^\circ$
\begin{equation}
\Omega< 40^\circ \quad \mbox{\&} \quad \Omega > 331^\circ \qquad (68\;{\rm per\;cent}\;{\rm C.L.})
\end{equation}
and
\begin{equation}
\Omega< 74^\circ \quad \mbox{\&} \quad \Omega > 297^\circ \qquad (95\;{\rm per\;cent}\;{\rm C.L.}) .
\end{equation}

%%%%%%%%%%

\subsection{Orbital period variations}
\label{subsec:period}

There are several effects that can contribute to changes in the observed
orbital period of a binary system that can be either intrinsic to the orbit or
just kinematic effects. The most important terms are:
\begin{equation}
\label{eq:period}
\dot{P_{b}} = \dot{P_{b}}^{\dot{m}} + \dot{P_{b}}^{T} + \dot{P_{b}}^{D} + 
              \dot{P_{b}}^{GW} + \dot{P_{b}}^{\dot{G}}.
\end{equation}
$\dot{P_{b}}$, measured here for the first time $\dot{P_{b}} =
5.0(1.4) \times 10^{-14}$, is the observable rate of change of the
orbital period. The first and second terms, $\dot{P_{b}}^{\dot{m}}$,
$\dot{P_{b}}^{T}$, are the contributions from the mass loss from the
binary and from tidal torques respectively. They can both be neglected
in the case of \psr\ because the first one is very small, as shown
before $(\sim 10^{-16})$, and the second is also small due to the lack
of interaction between the pulsar and the companion.

The third term, $\dot{P_{b}}^{D}$, is identical to the first term of
equation (\ref{eq:dotA1}).  In order to account for the Galactic
acceleration we have extended the \cite{dt91} expression (for a flat
rotation curve) to high Galactic latitudes
\begin{eqnarray}
\label{eq:Galactic}
  \left(\frac{{\dot P}_b}{P_b}\right)^{Gal} &=& -\frac{K_z|\sin b|}{c}
  \nonumber\\ &&  
  - \frac{\Omega_{\odot}^2 R_{\odot}}{c}\left(\cos l + \frac{\beta}{\beta^2 
  + \sin^2l}\right) \cos b \:, 
\end{eqnarray}
where $\beta \equiv (d/R_{\odot})\cos b - \cos l$. $K_z$ is the vertical
component of Galactic acceleration taken from \cite{hf04b}, which for Galactic
heights $z \equiv |d\sin b| \le 1.5$ kpc can be approximated with sufficient
accuracy by 
\begin{equation}
  K_z (10^{-9}\,{\rm cm\,s^{-2}}) \simeq 2.27\,z_{\rm kpc}  
                 + 3.68\left(1 - e^{-4.31\,z_{\rm kpc}} \right) \:,
\end{equation}
where $z_{\rm kpc} \equiv z({\rm kpc})$. $R_{\odot} = 8.0 \pm 0.4$
\citep{esg+03} and $\Omega_{\odot} = 27.2 \pm 0.9$\,km\,s$^{-1}$\,kpc$^{-1}$
\citep{fw97} are the Sun's Galactocentric distance and Galactic angular
velocity (= Oort's $A-B$). For the pulsar's Galactic coordinates of $l =
160.3^\circ$ and $b = 50.9^\circ$ we find
\begin{equation}
  {\dot P}_b^{\rm Gal} = (-5.6 \pm 0.2) \times 10^{-15} \;.
\end{equation}

We also calculate the contribution due to the Shklovskii effect
according to the following:
\begin{equation}
\label{eq:shk}
\dot{P_{b}}^{Shk} = \frac{(\mu_{\alpha}^{2} + \mu_{\delta}^{2})d}{c}P_{b}
                  = (7.0 \pm 0.7) \times 10^{-14} ,
\end{equation}
where we used the measured proper motion and the weighted mean of the
distance discussed earlier, $d$. So, by summing we yield the Doppler correction:
\begin{equation}
\label{eq:doppler}
\dot{P}_{b}^{D} = \dot{P}_{b}^{Gal} + \dot{P_{b}}^{Shk} 
                = (6.4 \pm 0.7)\times 10^{-14} .
\end{equation}

The fourth term, $\dot{P_{b}}^{GW}$, is the contribution due to
gravitational wave emission. In general relativity, for circular
orbits it is given by
\begin{equation}
\label{eq:gr}
\dot{P_{b}}^{GW}=\dot{P_{b}}^{GR} = -\frac{192\,\pi}{5}
  \left(\frac{2\pi}{P_b}\right)^{5/3}\frac{(T_{\odot}\,m_{c})^{5/3}q}{(q+1)^{1/3}}.
\end{equation}
For \psr\ we find $\dot{P_{b}}^{GW} = (-1.1 \pm 0.2) \times 10^{-14}$.  

All the previous terms are the ones that are expected to contribute by
using GR as our theory of gravity. However, most alternative theories
of gravity predict an extra contribution to the observed orbital
period variation, via dipole radiation (see \cite{wil93, wil01} and
references therein). This dipolar gravitational radiation results
from the difference in gravitational binding energy of the two bodies
of a binary system, and is expected to be much larger than the
quadrupolar contribution, especially if the binding energies of the
two bodies of the binary system differ significantly.  Thus, the
case of \psr\, , where there is a pulsar-WD system, is ideal for testing
the strength of such emission. One finds for small-eccentricity
systems
\begin{equation}
\label{eq:dipole1}
\dot{P_{b}}^{dipole} = -4\pi^2 \, \frac{T_\odot \mu}{P_b} \, \kappa_{D} {\cal S}^{2} \:, 
\end{equation}
where $m_c$ is expressed in units of solar masses. $\kappa_{D}$ refers to the
dipole self-gravitational contribution, which takes different values for
different theories of gravity (zero for GR) and ${\cal S} = s_p - s_c$ is the
difference in the ``sensitivities'' of the two bodies (see \cite{wil93} for
definition), and $\mu$ is the reduced mass, $m_{p}m_{c}/M$, of the system. The
sensitivity of a body is related to its gravitational self-energy
$\varepsilon$. In the post-Newtonian limit $s \simeq \varepsilon /mc^2$, which
gives $\sim 10^{-4}$ for a white dwarf \citep{wil01}. Hence, we can neglect
$s_c$ in equation (\ref{eq:dipole1}) since $s_p \sim 0.2$. Using the mass
ratio $q$, equation (\ref{eq:dipole1}) can be written as
\begin{equation}
\label{eq:dipole2}
\dot{P}_b^{dipole} = -4\pi^2 \, \frac{T_{\odot} m_c}{P_b} \,
                     \frac{q}{q+1}\,\kappa_D s_p^2 \:.
\end{equation}
For a specific theory of gravity $\kappa_D$ is known and $s_p$ can be
calculated as a function of the equation-of-state of neutron
star-matter. 

Finally, there are theories that predict that the locally measured
gravitational constant $G$ changes with time as the universe expands. A
changing gravitational constant would cause a change in the orbital period,
which for neutron star-WD systems can be written as
\begin{equation}
\label{eq:G}
\dot{P}_b^{\dot{G}} = -2\,\frac{\dot{G}}{G} 
                      \left[1 - \left(1+\frac{m_c}{2M}\right) s_p\right] P_{b} \:
\end{equation}
\citep{dgt88, nor90}.

The intrinsic change of the orbital period is the observed value minus
the Doppler correction term from equation (\ref{eq:doppler}):
\begin{equation}
\label{eq:pbdotintr}
\dot{P}_b^{intr} = \dot{P}_b - \dot{P}_b^D = (-1.5 \pm 1.5) \times 10^{-14} \:,
\end{equation}
which agrees well with the GR prediction given above as
\begin{equation}
\label{eq:pbdotzero}
\dot{P}_b^{exc} = \dot{P}_b^{intr} - \dot{P}_b^{GR} = (-0.4 \pm 1.6) \times 10^{-14} \:.
\end{equation}
Hence, there is no need for a $\dot{P}_b^{dipole}$ or $\dot{P}_b^{\dot{G}}$ to
explain the observed variation of the orbital period. On the other hand, this
can be used to set limits for a wide class of alternative theories of gravity,
which we will show in the following sections.

%%%%%%%%%%

\subsection{A generic limit for dipole radiation}

A tight system comprising a strongly self-gravitating neutron star and
a weakly self-gravitating white dwarf should be a very efficient
emitter of gravitational dipole radiation, if there is any deviation
from general relativity that leads to a non-vanishing $\kappa_D$ in
equation (\ref{eq:dipole1}). Hence, observations of such systems are
ideal to constrain deviations of that kind. \psr\ turns out to be a
particularly useful system to conduct such a test, since: (1) the
white-dwarf nature of the companion is affirmed optically, (2) the
mass estimates in this double-line system are free of any explicit
strong-field effects\footnote{The mass estimation for the weakly self
  gravitating white dwarf companion is done with Newtonian gravity
  \citep{cgk98}, and in any Lorentz-invariant theory of gravity the
  theoretical prediction for the mass ratio does not contain any
  explicit strong-field-gravity effects \citep{dam07}.}, which are a
priori unknown, if we do not want to restrict our analysis to specific
theories of gravity, (3) the estimated mass of the pulsar seems to be
rather high, which is important in the case of strong field effects that
occur only above a certain critical mass, like the spontaneous
scalarisation \citep{de93}.

In the previous section we have shown that the change in the orbital
period is in full agreement with the prediction by general relativity,
once the kinematic contributions are accounted for. Hence, any
deviations from general relativity leading to a different
$\dot{P}_b^{GW}$ is either small or compensated for a potential
$\dot{P}_b^{\dot G}$. However, we can already limit the variation of the
gravitational constant by using the published limit of $\dot{G}/G
= (4 \pm 9) \times 10^{-13}$\,yr$^{-1}$ from the Lunar Laser Ranging
(LLR) \citep{wtb04}. In combination with equation
(\ref{eq:G}) it gives $\dot{P}_{b}^{\dot{G}} = (-1 \pm 3) \times 10^{-15}$, for
the most conservative assumption $s_p = 0$. Hence, $\dot{P}_b^{dipole}
= (-0.2 \pm 1.6) \times 10^{-14}$, which with the help of equation
(\ref{eq:dipole2}) converts into
\begin{equation}
\label{eq:ap}
\kappa_D s_p^2 = (0.5 \pm 6.0)\times 10^{-5} \qquad (95\;{\rm per\;cent} \; {\rm C.L.}) \:.
\end{equation}
Furthermore, if we assume $s_p = 0.1 (m_p/M_\odot)$ (c.f.~\cite{de92}) we find
\begin{equation}
\label{eq:kappaD_LLR}
\kappa_D = (0.2 \pm 2.4) \times 10^{-3} \qquad (95\;{\rm per\;cent} \; {\rm C.L.}) \:. 
\end{equation}
This number improves upon the previously published limit for \psr\
\citep{lcw+01} by more than an order of magnitude.

For the tensor-scalar theories of \cite{de96} $\kappa_D {\cal S}^2 \simeq
(\alpha_p - \alpha_c)^2 \simeq (\alpha_p - \alpha_0)^2 < 6 \times 10^{-5}$,
assuming that the effective coupling strength of the companion WD to the
scalar field, $\alpha_c$, is much smaller than the pulsars and is
approximately $\alpha_0$, where $\alpha_0$ is a reference value of the
coupling at infinity. This value improves slightly on the previously published
limit of $7\times 10^{-5}$ \citep{nss+05}, obtained from PSR~J0751+1807. If
the non-linear coupling parameter $\beta_0$ is of order 10 or larger, then
neutron stars are much more weakly coupled to the scalar field than white
dwarfs \citep{esp04}. In this case, for \psr, $(\alpha_p - \alpha_c)^2 \approx
\alpha_0 < 6 \times 10^{-5}$, which is an order of magnitude weaker than the
limit $3.4 \times 10^{-6}$ from PSR~J1141$-$6545 \citep{bbv08}. Actually, in
tensor-scalar theories of gravity the latter is possibly the most constraining
pulsar binary system. However, since there has been no optical identification
of the companion, that could establish its WD nature without mass
determination based on a specific gravity theory, it is not yet possible to
derive a general theory independent limit for dipole radiation from
PSR~J1141$-$6545, as done here with \psr.

In the future, more accurate determination of the distance and improvement
of our $\dot{P}_b$ value, could further increase the precision of the
\psr\ limit.

%%%%%%%%%%

\subsection{Combined limits on $\dot{G}$ and the dipole radiation with
millisecond pulsars}
\label{subsec:Gdot}

In the previous section we have used the LLR limit for $\dot G$ in order to
provide a test for dipole radiation with a single binary pulsar system. On the
other hand, a generic test for $\dot G$ cannot be done with a single binary
pulsar, since in general theories that predict a variation of the
gravitational constant typically also predict the existence of dipole
radiation \citep{wil93}.\footnote{It is interesting to point out, that in the
  Jordan-Fierz-Brans-Dicke theory $\dot{P}_b^{dipole} + \dot{P}_b^{\dot G} =
  0$ for binary pulsars with white-dwarf companions that have orbital periods
  $\sim$10 days.}  From equations (\ref{eq:dipole1}) and (\ref{eq:G}) we can
see that ${\dot P}_b^{\dot G} \propto P_b$ whereas ${\dot P}_b^{\rm dipole}
\propto P_b^{-1}$. Hence, one can combine any two binary pulsars, with tight
limits for ${\dot P}_b$ and different orbital periods, in a joint analysis to
break this degeneracy, and to provide a test for $\dot G$ and the dipole
radiation that is based purely on pulsar data. A formally consistent way of
doing this with white-dwarf binary pulsars is the application of equation
%
%\begin{eqnarray}
%  && \hspace{-7.5mm}
%  \frac{\dot{P}_b^{intr} - \dot{P}_b^{GR}}{P_b} =  
%    - 2 \frac{\dot G}{G} \left[1 - \left(1+\frac{m_c}{2M}\right) s_p\right]
%    - 4 \pi^2 \frac{T_\odot\mu}{P_b^2} \, \kappa_{\rm D} s_p^2 \nonumber\\ &&
%\end{eqnarray}
%
\begin{equation}
 \frac{\dot{P}_b^{exc}}{P_b} =  
    - 2 \frac{\dot G}{G} \left[1 - \left(1+\frac{m_c}{2M}\right) s_p\right]
    - 4 \pi^2 \frac{T_\odot\mu}{P_b^2} \, \kappa_{\rm D} s_p^2
\end{equation}
(see equations (\ref{eq:G}) and (\ref{eq:dipole1})) to both binary pulsars,
and solving in a Monte-Carlo simulation this set of two equations for $\dot
G/G$ and $\kappa_{\rm D}$. This procedure properly accounts for the
correlations due to this mutual dependence, and thus provides a self
consistent test for $\dot G$ and the dipole radiation, that does not rely on
LLR limits or theory specific assumptions. There remains the problem of
getting a good estimate for $s_p$ in a general theory independent test. As
before, we will use $s_p = 0.1 \, (m_p/M_\odot)$ keeping in mind that the
limits given below are subject to certain changes, if a different assumption
for $s_p$ is made.

With its short orbital period and its fairly well determined masses \psr\ is
an ideal candidate for such a combined analysis. Presently, the best binary
pulsar limit for $\dot G$ comes from PSR~J0437$-$4715 \citep{vbv+08, dvtb08},
where ${\dot P}_b^{\rm dipole}=0$ has been used in the analysis to obtain the
limit for $\dot G$. Using this pulsar in combination with \psr\ in a joint
analysis as introduced above gives, with a 95 per cent C.L.,
\begin{equation}
  \frac{\dot G}{G} = (-0.7 \pm 3.3) \times 10^{-12} \; {\rm yr}^{-1} 
                   = (-0.009 \pm 0.045) \, H_0
\end{equation}
and
\begin{equation}
  \kappa_{\rm D} = (0.3 \pm 2.5) \times 10^{-3}
\end{equation}
where $H_0 = 74$\,km\,s$^{-1}$\,Mpc$^{-1}$ has been used as a value for the
Hubble constant \citep{rmc+09}. Our pulsar test therefore restricts $\dot G/G$ 
to less than a 20th of the expansion rate of the Universe.

\begin{figure}
\begin{center}
    \mbox{\includegraphics[width=0.48\textwidth]{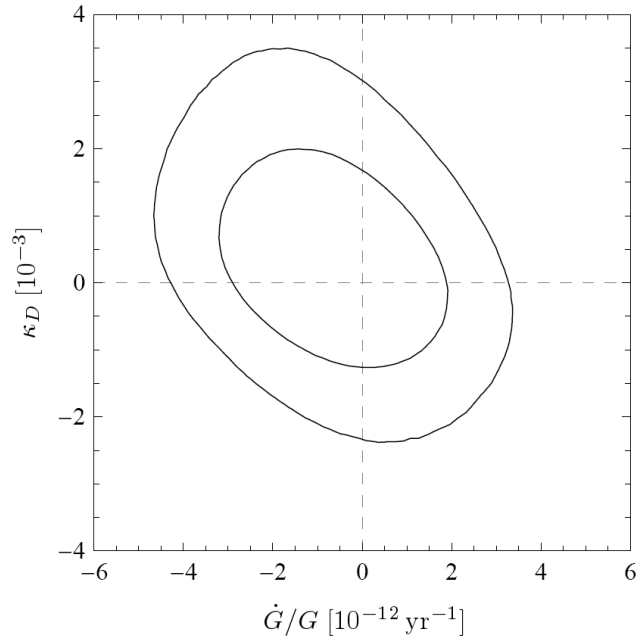}}
\end{center}
\caption{Contour plots of the one and two $\sigma$ confidence regions on $\dot
  G/G$ and $\kappa_D$ jointly. The elongation of the regions reflects the
  correlation due to the mutual dependence of the two systems, \psr\ and
  PSR~J0437$-$4715, in this combined test.}
\label{fig:monteKG}
\end{figure}

The limit for $\dot G$ given here is clearly weaker than the one given in
\cite{dvtb08}. The main reason for this is that the equation for
$\dot{P}_b^{\dot G}$ used by \cite{dvtb08} does not account for the
sensitivity of the pulsar as in equation (\ref{eq:G}).  Furthermore, the
combined analysis still allows for a certain range for ${\dot P}_b^{\rm
  dipole}$ in PSR~J0437$-$4715, leading to a somewhat weaker limit compared
with an analysis that uses ${\dot P}_b^{\rm dipole} = 0$, as can be seen in
Fig. \ref{fig:monteKG}. Although this limit for $\dot G$ is weaker than the
LLR limit, it still provides a useful independent addition to the LLR
result, as has been argued in \cite{vbv+08}.

%-------------------------------------------------------------------------------

%-------------------------------------------------------------------------------

The limit for the dipole radiation is slightly weaker than the one given in
the previous section. However, in contrast to the limit of the previous
section, the limit here does not rely on the LLR result for $\dot G$, and
therefore constitutes an independent test based solely on binary pulsar
observations.

We would like to stress two facts about the advantage of combining
specifically these two binary pulsars.  Firstly, in both cases the companion
white dwarf is identified optically, and its non-compact nature is ascertained
independently of the underlying theory of gravity.  Secondly, the two pulsars
seem to be rather heavy and similar in mass ($\sim 1.7 M_\odot$)\footnote{In
  general, PSR~J0437$-$4715 does not allow the determination of the pulsar
  mass, since this requires the mass function, which contains explicit
  strong-field contributions. Within the generic class of conservative gravity
  theories (\cite{wil93, dt92}), for instance, only the effective
  gravitational mass, ${\cal G}m_p$, of PSR~J0437$-$4715 can be
  determined. However, if one assumes that ${\cal G}$ deviates less than 20
  per cent from $G$, the pulsar mass is in the range of 1.5 to 2.0 solar
  masses.}, which is important in case we have effects like spontaneous
scalarisation above a critical neutron star mass \citep{de93}. In the future,
more accurate measurements of $\dot{P}_{b}$ and distance of the two pulsars
could constrain even more our derived limits.

%%%%%%%%%%%%%%%%%%%%%%%%%%%%%%%%%%%%%%%%%%%%%%%%%%%%%%%%%%%%%%%%%%%%%%%%%%%%%%%%

\section{Conclusions}
\label{sec:Concl}

We have presented results from the high precision timing analysis of 15\,yr
of EPTA data for \psr. A first ever measurement of the timing parallax
$\pi = 1.2(3)$ and distance has been obtained for this pulsar. Combined with
information from optical observations of the WD companion an improved 3D velocity
has been derived for the system. This information enables the
derivation of the complete evolutionary path of the pulsar in the Galaxy,
showing that it spent most of its lifetime far away from the solar system
orbit. In addition, an improved limit on the extremely low intrinsic
eccentricity, $e < 8.4 \times 10^{-7}$ (95 per cent C.L.), has been acquired, which
agrees well with the theoretical eccentricity-orbital period relation \citep{pk94}.

Of particular interest is the measurement of the variation of the
projected semi-major axis, $\dot{x} = 2.3(8) \times 10^{-15}$ which is
caused by a change in the orbital inclination as the system moves
relative to the SSB. This measurement allowed us to set limits on the
positional angle of the ascending node, for the first time, the last
unknown parameter in fully describing the orientation of this binary
system.

As a result of the significant measurement of the change in the orbital period
of the system, $5.0(1.4) \times 10^{-14}$, and the identified nature of the
two bodies in this binary system, tests for alternative gravity theories could
be performed. Firstly, a stringent, generic limit for the dipole radiation
has been obtained from \psr, $\kappa_D s_p^2 = 0.5 \pm 6.0 \times 10^{-5}$
(95 per cent {\rm C.L.}), with the use of the $\dot{G}$ limit from LLR. Secondly, in
a self consistent analysis we have used \psr\ together with PSR J0437$-$4715
to derive a combined limit on the dipole radiation and the variation of the
gravitational constant, $\kappa_D = (0.4 \pm 2.6) \times 10^{-3}$ and $\dot G
/ G = (-0.7 \pm 3.3) \times 10^{-12} \; {\rm yr}^{-1} $ (95 per cent {\rm C.L.})
respectively. These limits have been derived just with the use of millisecond
pulsar-WD binaries and are valid for a wide class of alternative theories of
gravity.

%%%%%%%%%%%%%%%%%%%%%%%%%%%%%%%%%%%%%%%%%%%%%%%%%%%%%%%%%%%%%%%%%%%%%%%%%%%%%%%%

\section*{Acknowledgements}

We are very grateful to all staff at the Effelsberg, Westerbork, Jodrell Bank
and Nan\c cay radio telescopes for their help with the observations. Kosmas
Lazaridis was supported for this research through a stipend from the
International Max Planck Research School (IMPRS) for Astronomy and
Astrophysics at the Universities of Bonn and Cologne. We are grateful to 
Paulo Freire for valuable discussions.

%\bibliographystyle{mn2e}
%\bibliography{lazaridis} 
%\bibliography{journals}

\end{document}